# Resistive Switching Behaviour of Organic Molecules

Bapi Dey, Surajit Sarkar, Hritinava Banik, Syed Arshad Hussain*

*Thin film and Nanoscience Laboratory, Department of Physics, Tripura University, Suryamaninagar*
*\* Corresponding author*
*Email: sa_h153@hotmail.com, sahussain@tripurauniv.in*
*Ph: +919402122510 (M), +91381 2379119 (O)*
*Fax: +913812374802 (O)*

**Abstract**

Organic electronics is very promising due to the flexibility, modifiability as well as variety of the available organic molecules. Efforts are going on to use organic materials for the realization of memory devices. In this regard resistive switching devices surely will play a key role. In this paper an effort has been made to illustrate the general information about resistive switching devices as well as switching mechanisms involving organic materials. As a whole a general overview of the emerging topic resistive switching has been given.

## 1. Introduction

Over the decades, organic electronics has attracted much attention due to the distinctive advantages of organic materials, such as low cost of processing, lightweight, mechanical flexibility, ambient processing, printability etc. Their exceptional electrical performances have made them a suitable candidate for memory technologies as well as other various types of electronic and optoelectronic devices, such as diodes, sensors, and organic light-emitting diodes (OLED), organic field effect transistors (OFET), solar cells, lasers, detectors etc where, organic materials have been used as the active medium [1-7]. Also conducting polymers have gained serious attention into inorganic dominated electronic and opto-electronic devices, some of which have already attained commercial viability. Data-storage and switching applications of polymer-based devices are being investigated for quite some times now. However, the same using organic materials have been lacking.

As the industry moves from bulk to molecular electronics, there is a growing trend to revisit voltage-induced switching phenomena in conjugated organics, which was initially observed more than thirty years ago [8-9]. The basic component of logic elements is a switch. Conductance switching is based on the study of single molecules or single molecular layer has attracted great attention in recent years due to their potential future applications in memory / storage systems [10]. Switching behaviour of organic molecules have made them an essential candidate for making organic resistive memory devices in which active organic materials possess at least two stable resistance states. Such devices can be switched between a high resistance (OFF) and a low resistance (ON) state by suitable applied voltage [11]. With respect to device fabrication they are very much advantageous due to simple device structures, low fabrication costs, and printability. Furthermore, their exceptional electrical performances such as a nondestructive reading process, nonvolatility, a high ON/OFF ratio, and a fast switching speed meet the requirements for viable memory technologies. Depending on the molecular structure, conformation, and types of contacts, a wide range of switching behaviours has already been observed in current–voltage (I–V) characteristics for metal/molecule/metal structures [10]. These types of organic switching devices are the promising candidates for next generation information storage and future optoelectronic devices [12-13]. Apart from switching and memory applications, use of such switching devices in logic elements of integrated circuits are also feasible. Of late study of resistive switching have been attracted much attention from researchers as well as from application oriented communities as observed from the increasing number of research papers per year as shown in figure 1.

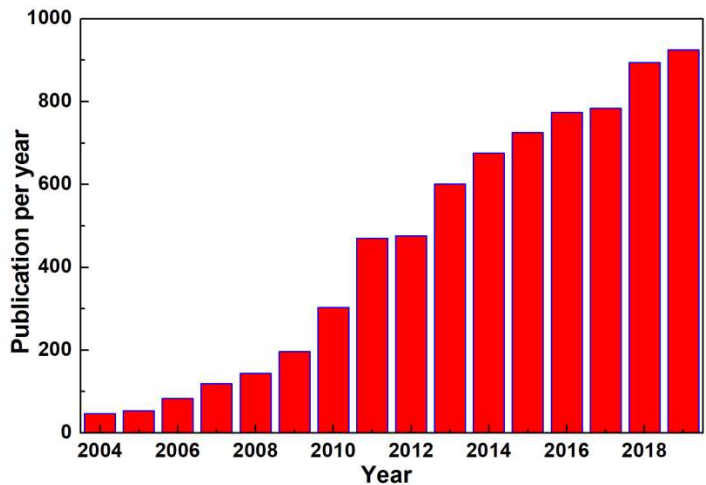

Fig.1. Publications per year from 2004 to 2019. Data collected from web of Science site (http://apps.webofknowledge.com/) using the search topic = 'Resistive Switching'.

A number of mechanism and strategies responsible for the two switching states in a variety of organic molecules have been proposed such as conformational change, rotation of functional groups, charge transfer, filamentary conduction, space charge and traps, ionic conduction and oxidation-reduction process etc [13-17]. There are several reports with useful insights into scientific and technical issues related to resistive switching phenomenon in organic molecules [7-13]. However, full understanding of the underlying physics behind resistive switching behaviour observed in organic systems is still lacking. This feature article illustrates the general information about resistive switching devices as well as switching mechanisms involving organic materials and will give a general overview of the emerging topic resistive switching.

## 2. Classification of resistive switching phenomena

Depending on the ability to retain information, the resistive switching phenomenon using organic molecules are categorized into two classes, –
(1) Volatile threshold switching. and
(2) Nonvolatile memory switching

In case of nonvolatile memory switching after the removal of external voltage both the low resistance state (LRS) and high resistance state (HRS) can be retained, whereas, in threshold switching only the HRS is stable at low applied voltages [18].

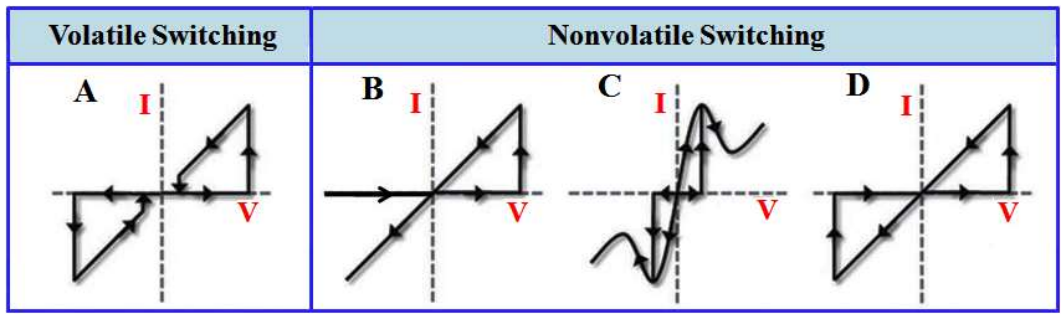

Fig.2. Schematic of the I-V characteristics for the volatile and non-volatile switching devices - A) volatile Threshold switching, B) WORM switching, C) unipolar switching and D) bipolar switching behaviour [reprinted with permission from ref. 19].

## 2.1. Volatile threshold switching

Volatile threshold switching describes resistive switching phenomena for which there is only one stable state with no external bias [20-22]. Similar to dynamic random access memory (DRAM), volatile switching requires periodic refreshing as a result of the loss of stored information. [23-24]. Figure 2A shows a typical I–V curve for volatile threshold switching. At $V_{threshold}$, the device changes from the HRS to the LRS. The LRS is stable over only a certain range of applied biases, and when the applied bias falls below this range, the device returns to the HRS. Threshold switching have found many potential technological applications including electrical switches [25], smart windows [26], terahertz nanoantennas [27], and memory metamaterials [28]. In addition, threshold switching can also be used as a selector to solve a sneak path problem occurring in a RRAM crossbar array [25, 29].

## 2.2. Non-volatile memory switching

Nonvolatile switching devices are electrically programmed organic memory device which can retain data for a long period of time. This is similar to conventional flash memory. Nonvolatile switching is often classified into three types based on current-voltage (I – V) characteristics: write-once-read-many-times (WORM), unipolar, and bipolar switching memory (B, C, and D in Fig. 2).

### 2.2.1. WORM memory

WORM-type memory devices show electrically irreversible switching characteristics, and the original state is never recovered [30]. Once the devices are set to LRS, they do not switch back to HRS, even after the application of large positive or negative voltages (Fig. 2B). As only unidirectional switching from HRS to LRS is observed when either a positive or negative sweep was applied, the resistance states of these devices were electrically irreversible. Thus, information once written in this type of memory device cannot be reprogrammed but can be read for a long period of time. Hence, the memory devices are categorized as write-once-read-many (WORM) type, which is suitable for non-editable database, archival memory, electronic voting and radio frequency identification applications etc.

### 2.2.2. Unipolar switching

Unipolar memory systems exhibit electrically reversible switching. Unipolar memory devices use the same voltage polarity to write and erase [31]. Figure 2C shows a typical I–V characteristic for unipolar switching. If the device is initially in the LRS and a positive external bias voltage $V_{ext}$ (>0) is applied to it, when the voltage $V_{reset}$ is reached, the resistance abruptly increases and the device enters the HRS. After that if again a positive $V_{ext}$ is applied to the device in the HRS, an abrupt reduction in resistance occurs at $V_{set}$, and the device goes into the LRS again (Fig.2C). The set voltage ($V_{set}$) is always higher than the voltage at which reset ($V_{reset}$) occurred. In unipolar switching, the I–V curve is symmetric for both forward and reverse bias. When revese bias is applied, similar reset and set processes also occur. Therefore, external bias voltage with only one polarity are sufficient in real device operations. This is why such RS is called 'unipolar'.

### 2.2.3. Bipolar switching

Bipolar memory devices require both positive and negative voltage polarities [32]. The resistance state of this kind of memory devices are also electrically reversible. Once information is written in this type of memory device, it can be erased and the device is rewriteable. A typical I–V characteristic for bipolar switching is shown in Fig. 2D. Here a negative bias is required for the reset process, which takes the device from the LRS into the HRS, and a positive bias is required for the set process, which takes the device from the HRS into the LRS. As both polarities of the external bias voltage are required, this type of RS is called bipolar switching. Main difference between unipolar and bipolar switching devices is that in bipolar switching set and reset require opposite voltage polarities, whereas, in unipolar switching set and reset occurred at the same voltage polarity. In bipolar switching the switching processes are field driven and in unipolar switching at least one of the processes (either set or reset) is not field driven but temperature/diffusion driven.

## 3. Device structure

Resistive switching devices are usually composed of an active layer of molecules sandwiched between two electrodes. The active layer in such devices is usually a thin film of the selected materials. Different conventional thin film forming techniques can be used to prepare such layers [7, 33]. A number of molecules (both inorganic and organic as well as their mixture) have already been tested [14-33]. Inorganic materials have advantages over organic one with respect to switching stability, while organic molecules stand out in terms of high mechanical flexibility, ease of fabrication process and cost effectiveness. Selection of suitable electrode material is also very crucial as they often affect the switching behaviour [34].

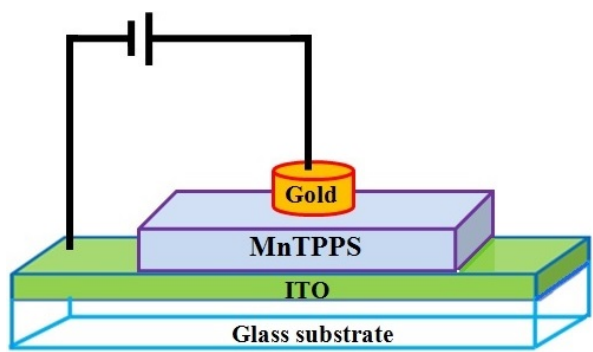

Fig. 3. Schematic representation of a switching device having configuration with active layer of MnTPPS and electrodes – ITO and gold [reprinted with permission from ref. 33].

## 4. Organic materials suitable for optoelectronic device application

Various organic materials showed switching characteristics based on changes in resistance upon application of biasing. A range of materials such as small organic molecules, polymers and composite materials sandwiched between two electrodes may be used in the switching devices. Over the last few decades, tremendous progress has been made in the development and investigation of new organic materials [35−36] with a readily polarizable structure for the fabrication of novel devices. A conjugated π-system end-capped with a strong electron donor group, such as −NR$_2$ and OR, and a strong electron acceptor group, such as −NO$_2$, cyano (CN), or imine (C=NR), in a molecule generates a dipolar push−pull system (D−π−A) that assures intramolecular charge transfer and low energy barrier and shows an intense charge-transfer band. The extent of electronic or optoelectronic behaviour of such organic materials depends primarily on their chemical structure, high chemical and thermal robustness, good solubility in common organic solvents, and availability in reasonable quantities. Also various five- and six-membered heterocycles can be utilized as suitable π-conjugated chromophore backbones because they act as auxiliary donors or acceptors and improve the overall polarizability of the chromophore [37−38]. Functional groups that enhances the reduction process in the organic system is also very crucial.

## 5. Mechanisms involved in switching behaviour

Full understanding of the mechanism and underlying physics behind the switching phenomenon in the devices based on organic molecules are yet to fully understand. The electrical conduction mechanism in organic molecules has been renewed time to time. A substantial amount of research has been dedicated to the understanding of the switching phenomena.

A number of mechanisms and strategies responsible for the two switching states in a variety of organic molecules have been proposed, such as filamentary conduction, space charge and traps, charge transfer, conformational changing, ionic conduction, rotation of functional group, and reduction–oxidation process, electron tunnelling and hopping etc [14–17].

Few of these proposed switching mechanisms have been discussed bellow.

*5.1. Filamentary conduction*

Filamentary conduction occurs via two types of filament paths when the flow of current is limited to highly localized regions in a junction area. One is associated with the carbon-rich filaments formed due to local degradation of organic films and the other is related to the metallic bridges that result from the migration of electrodes. This type of switching mechanism is often used by the researchers to explain resistive switching phenomena observed in a variety of organic memory devices. For example, penetration of Cu ions into organic layer under forward bias could result in filament formation in Cu/poly(3-hexylthiophene) (P3HT)/Al device, which was experimentally proved by secondary ion mass spectroscopy analysis [39]. Similar metallic filaments could also be formed in poly(ethylenedioxythiphene):poly(styrenesulfonate) (PEDOT:PSS) or poly(4-hydroxystyrene) composite materials that contain homogeneously dispersed Au clusters [40]. The stability and irreversibility of highly conductive percolation paths in a polymer resulted WORM – type memory [40]. On the other hand electrical evidence of filamentary conduction was reported in terms of metallic and area independent behaviours of high conductance state [41-42]. As the filamentary conduction system is highly localized process, it offers the chance of creating nanoscale memory devices.

*5.2. Space charge and traps*

In case of ohmic contact between the electrodes and the organic material and for trap free insulator, accumulation of charge carriers near the electrode builds up a space charge. As a result mutual repulsion between individual charges occurs that restricts the total charge injected into a sample and the resulting current is called space charge limited current (SCLC). Several factors responsible for this phenomenon such as the injection of electrons or holes from the electrode, the presence of ionized dopants in interfacial depletion regions, and the accumulation of mobile ions at electrode interfaces. However, traps present in the bulk of the material or at interfaces reduce carrier mobility. When the traps are located at interfaces, they can affect the injection of charges into a material. Reports are there regarding the electrical bistability of some organic materials due to space charges [43]. Bistable switching in the devices prepared by organic materials arises from the accumulation of space charges. This space charges accumulate at the metal–polymer interface, there by restricts the electrical field that limits the injection of further charges into the organic layer [44]. The stored charges control charge injection and lead to hysteresis in the I–V curve. Switching behavior of trilayer structure composed of granular metals sandwiched between two organic layers have been explained in term of charge storage mechanism [45].

*5.3. Charge transfer*

Charge transfer is another important mechanism responsible for electrical switching behaviour of organic molecules in the ultrathin film. Charge transfer mainly occurs in the organic materials consisting of electron donor–acceptor systems in which electrical charges are partially transferred from the donor to the acceptor moiety [46]. Field-induced transfers are expected to occur most frequently in the charge transfer complexes. As a result a sharp increase in conductivity was observed after the charge transfer. Cu-tetracyanoquinodimethane (Cu-TCNQ) molecules showed switching characteristics due to charge transfer from Cu to TCNQ [46]. Molecules [6,6]-phenyl-C61-butyric acid methyl ester (PCBM) and tetrathiofulvalene (TTF) dispersed onto polystyrene (PS) matrix showed bi-stable resistance states due to electron transfer from the HOMO state of TTF to the LOMO state of PCBM [47].

*5.4. Conformational change*

Conformational change in the molecular backbone is another factor that governs the resistive switching mechanism in the organic molecules. This arises due to the effect of the applied electric field [48]. Rose Bengal (RB) molecules exhibits conductance switching in supramolecular matrices of polyelectrolytes due to conformational changes arising from the applied biasing [48]. In the low voltage region, the reverse bias induced electro-reduction of RB molecules facilitates the restoration of conjugation in the molecular backbone resulting in an ON state. However, at large applied bias two perpendicular planes present in RB molecules induce forward bias based conformational change and result in the conductance switching. The difference in the memory behaviour between PCz and PVBCz arose due to inherent differences in the degree of region-regularity and the ease of conformational relaxation caused by the packing density during device formation which hinder the extent of molecular conformational change [19]. Upon application of electric field randomly oriented carbazole groups in the organic molecules were rearranged into face to face orientations, facilitating carrier delocalization and transport resulting change in conduction states [48].

*5.5. Ionic conduction*

Ionic conduction usually occurs in polymers containing ionic groups. Compared to the activation energy required for electronic conduction, the movement of ions requires a relatively high activation energy, and a long transit time of the ions is also expected. These processes create a p-n junction and result in an asymmetrical I–V curve. Thus, rectifying bistable switching can be induced in organic systems using ionic conduction. Electrically rewritable switching effects have been observed in P3HT [49] and a sexithiophene poly (ethylene oxide) (6T-PEO) block copolymer [50]. Inorganic salts $LiCF_3SO_3$ and NaCl were used as dopants. It has been suggested that migration of the dopant ions into and out of the P3HT or 6T-PEO depletion layer at the aluminum Schottky contact causes the electrical bistability.

## 6. Switching behavious of organic molecules assembled onto Langmuir-Blodgett Films

Organic molecules assembled onto Langmuir-Blodgett (LB) films have been considered as the basic building blocks for next generation Molecular Electronics. Inherent control on the molecular organization in LB films and the ability to organize mono-molecular layer make this technique unparallel with respect to others [51]. Here, few examples of resistive switching of organic molecules assembled onto LB films have been mentioned. Imidazole derivatives tailored with various aromatic groups can form a donor−acceptor system and may show attractive properties in terms of conductivity and luminescence. It has been observed that imidazole derivative (2,4,5-triaryl imidazole) formed nanostructures when organized onto LB films and showed resistive switching behavior due to charge transfer between the donor−acceptor groups present in the molecule [52]. On the other hand imidazole derivative 1-benzyl-2,4,5-triaryl imidazole showed bipolar switching behavior, where, reduction – oxidation process played crucial rule for the observed bipolar switching [53]. These kind of bipolar switching of organic molecule is very important for memory application and logic element in integrated circuit. Switching devices suitable for write-once-read-many-times memory devices (WORM) and resistive-switching random access memory application (RRAM) based on amido-phenazine derivative – stearic acid mixtures have been demonstrated [7]. These kind of devices may have promising applications such as - non-editable database, archival memory, electronic voting, radio frequency identification etc. Both threshold as well bipolar resistive switching involving metalloporphyrin molecules assembled onto LB films have also been observed. Here two types of switching were observed simply by manipulating the device fabrication condition during LB film formation [33].

## 7. Conclusion

In conclusion it can be said that resistive switching devices can be used to implement memory devices as well as logic elements of integrated circuits involving organic materials and thus will make the organic electronics into reality. Researchers and scientists throughout the world are giving significant effort in this regard. However, lots of further research involving new organic materials under various conditions have been required. Collaboration of scientist from both basic sciences as well as applied sciences and engineering and industry are required. Keeping in mind about the potential of resistive switching, we sincerely hope that soon we will see organic materials as the building blocks of memory and logic elements in integrated circuits.

**Acknowledgements**

The authors are grateful to DST, for financial support to carry out this research work through FIST – DST project ref. SR/FST/PSI-191/2014. SAH is grateful to DST, for financial support to carry out this research work through DST, Govt. of India project ref. No. EMR/2014/000234. The authors are also grateful to UGC, Govt. of India for financial support to carry out this research work through financial assistance under UGC – SAP program 2016.

**References**

[1] M. Shao, X. Xu, J. Han, J. Zhao, W. Shi, X. Kong, M. Wei, D.G. Evans, X. Duan, Langmuir 27 (2011) 8233–8240.
[2] M.T. Blazquez, F.M. Muniz, S. Saez, Heterocycles 69 (2006) 73–81.
[3] G. Chen, C. Si, P. Zhang, B. Wei, J. Zhang, Z. Hong, H. Sasabe, J. Kido, Org. Electron. 51 (2017) 62–69.
[4] H. Jinno, T. Yokota, N. Matsuhisa, M. Kaltenbrunner, Y. Tachibana, T. Someya, Org. Electron. 40 (2017) 58–64.
[5] Y. Sun, L. Li, D. Wen, X. Bai, G. Li, Phys. Chem. Chem. Phys. 17 (2015) 17150–17158.
[6] X. Ban, K. Sun, Y. Sun, B. Huang, W. Jiang, Org. Electron. 33 (2016) 9–14.
[7] D.K. Maiti, S. Debnath, S.M. Nawaz, B. Dey, E. Dinda, D. Roy, S. Ray, A. Mallik, S.A. Hussain, Sci. Rep. 7 (2017) 13308–13317.
[8] S.R. Ovshinsky, Phys. Rev. Lett. 21 (1968) 1450–1453.
[9] A. Elsharkawi, J. Phys. Chem. Solids. 38 (1977) 95–96.
[10] J.M. Seminario, A.G. Zacarias, P.A. Derosa, J. Chem. Phys. 116 (2002).1671-1683.
[11] R. Waser, M. Aono, Nat. Mater. 6 (2007) 833-840.
[12] G. Liu, Q.D. Ling, E.T. Kang, K.G. Neoh, D.J. Liaw, F.C. Chang, C.X. Zhu, D.S.H. Chan, J. Appl. Phys. 102 (2007) 024502-024509.
[13] A. Bandyopadhyay, A. J. Pal, Chem. Phys. Lett. 371 (2003) 86-90.
[14] A. Sawa, Mater. Today 11 (2008) 28-36.
[15] H. Yildirim, R. Pachter, ACS Appl. Mater. Interfaces. 10 (2018) 9802-9816.
[16] T. Kondo, S.M. Lee, M. Malicki, B. Domercq, S.R. Marder, B. Kippelen, Adv. Funct. Mater. 18 (2008) 1112–1118.
[17] I. Hwang, M.J. Lee, G.H. Buh, J. Bae, J. Choi, J.S. Kim, S. Hong, Y.S. Kim, I.S. Byun, S.W. Lee, S.E. Ahn, B.S. Kang, S.O. Kang, B.H. Park, Appl. Phys. Lett. 97 (2010) 052106- 052108.
[18] S.H. Chang, J.S. Lee, D.W. Kim, T.W. Noh, Phys. Rev. Lett. 102 (2009) 026801-026804.
[19] B. Cho, S. Song, Y. Ji, T.W. Kim, T. Lee, Adv. Funct. Mater. 21 (2011) 2806–2829.
[20] G. Dearnale, A.M. Stoneham, D.V. Morgan, Rep. Prog. Phys. 33 (1970) 1129-1191.
[21] S. Seo, M.J. Lee, D.H. Seo, E.J. Jeoung, D.S. Suh, Y.S. Joung, I.K. Yoo, I.R. Hwang, S.H. Kim, I.S. Byun, J.S. Kim, J.S. Choi, B.H. Park, Appl. Phys. Lett. 85 (2004) 5655-5657.
[22] D. Adler, H.K. Henisch, N. Mott, Rev. Mod. Phys. 50 (1978) 209-220.
[23] T.J. Lee, S. Park, S.G. Hahm, D.M. Kim, K. Kim, J. Kim, W. Kwon, Y. Kim, T. Chang, M. Ree, J. Phys. Chem. C 113 (2009) 3855-3861.
[24] X. Zhu, W. Su, Y. Liu, B. Hu, L. Pan, W. Lu, J. Zhang, R.W. Li, Adv. Mater. 24 (2012) 3941–3946.
[25] S. H. Chang, S. B. Lee, D. Y. Jeon, S. J. Park, G. T. Kim, S. M. Yang, S. C. Chae, H. K. Yoo, B. S. Kang, M.-J. Lee, and T. W. Noh, Adv. Mater. 23 (2011) 4063-4067.
[26] K. Kato, P.K. Song, H. Odaka, Y. Shigesato, Jpn. J. Appl. Phys. Part 1. 42 (2003) 6523-6531.
[27] M. Seo, J. Kyoung, H. Park, S. Koo, H.-S. Kim, H. Bernien, B. J. Kim, J. H. Choe, Y.H. Ahn, H.T. Kim, N. Park, Q.H. Park, K. Ahn, D.-s. Kim, Nano Lett. 10 (2010) 2064-2068.
[28] T. Driscoll, H.T. Kim, B.G. Chae, B.J. Kim, Y.W. Lee, N.M. Jokerst, S. Palit, D.R. Smith, M.Di Ventra, D.N. Basov, Science 325 (2009) 1518-1521.
[29] M.J. Lee, Y. Park, D.S. Suh, E.H. Lee, S. Seo, D.C. Kim, R. Jung, B.S. Kang,S.E. Ahn, C.B. Lee, D.H. Seo, Y.K. Cha, I.K. Yoo, J.S. Kim, B.H. Park, Adv. Mater. 19 (2007) 3919-3923.
[30] J. Lin, D. Ma, J. Appl. Phys. 103 (2008) 024507-024510.
[31] B. Cho, T.W. Kim, M. Choe, G. Wang, S. Song, T. Lee, Org. Electron. 10 (2009) 473-477.
[32] J. Ouyang, Y. Yang, Appl. Phys. Lett. 96 (2010) 063506-063508.
[33] B. Dey, S. Chakraborty, S. Chakraborty, D. Bhattacharjee, Inamuddin, A. Khan, S.A. Hussain Org. Electron. 55 (2018) 50 – 62.
[34] S. Gao, C. Song, C. Chen, F. Zeng, F. Pan, J. Phys. Chem. C. 116 (2012) 17955-17959.
[35] Y.F. Sun, Y.P. Cui, Dyes Pigm. 81 (2009) 27-34.
[36] S. Huang, Z. Li, S. Li, J. Yin, S. Liu, Dyes Pigm. 92, (2012) 961-966.
[37] J. Kulhánek, F.J. Bureš, J. Org. Chem. 8 (2012) 25-49.
[38] J. Yuan, Z. Li, M. Hu, S. Li, J. Yin, S.H. Liu, Photobiol. Sci. 10 (2011) 587-591.
[39] W.J. Joo, T.L. Choi, K.H. Lee, Y. Chung, J. Phys. Chem. B 111 (2007) 7756-60.
[40] S. Sivaramakrishnan, P.J. Chia, Y.C. Yeo, L.L. Chua, P.K.H. Ho, Nat. Mater. 6 (2007) 149-155.

[41] W.J. Joo, T.L. Choi, J. Lee, S.K. Lee, M.S. Jung, N. Kim, J.M. Kim, J. Phys. Chem. B 110 (2006) 23812-23816.
[42] T.W. Kim, H. Choi, S.H. Oh, M. Jo, G. Wang, B. Cho, D.Y. Kim, H. Hwang, T. Lee, Nanotechnology 20 (2009) 025201-025205.
[43] H.S. Majumdar, A. Bandyopadhyay, A. Bolognesi, A.J. Pal, J. Appl. Phys. 91 (2002) 2433-2437.
[44] S. Das, A.J. Pal, Appl. Phys. Lett. 76 (2000) 1770-1772.
[45] J.G. Simmons, R.R. Verderber, Proc. R. Soc. London, Ser. A 301 (1967) 77-102.
[46] R.S. Potember, T.O. Poehler, D.O. Cowan, Appl. Phys. Lett. 34 (1979) 405-407.
[47] C.W. Chu, J. Ouyang, J.H. Tseng, Y.Yang, Adv. Mater. 17 (2005) 1440-1443.
[48] A. Bandyopadhyay, A.J. Pal, Appl. Phys. Lett. 84 (2004) 999-1001.
[49] J.H.A. Smits, S.C.J. Meskers, R.A.J. Janssen, A.W. Marsman, D.M.de Leeuw, Adv. Mater. 17 (2005) 1169-1173.
[50] F. Verbakel, S.C.J. Meskers, R.A.J. Janssen, Chem. Mater. 18 (2006) 2707-2012.
[51] S.A. Hussain, B. Dey, D. Bhattacharjee, N. Mehta, Heliyon 4 (2018) e01038.
[52] B. Dey, P. Debnath, S. Chakraborty, B. Deb, D. Bhattacharjee, S. Majumdar, S.A. Hussain, Langmuir 33 (2017) 8383 – 8394.
[53] B. Dey, S. Suklabaidya, S. Majumdar, P.K. Paul, D. Bhattacharjee, S.A. Hussain, Chemistry Select 4 (2019) 9065–9073.